\documentclass[12pt]{article}

\newcommand{\bc}{\begin{center}}\newcommand{\ec}{\end
{center}}
\newcommand{\be}{\begin{equation}}\newcommand{\bit}
{\begin{itemize}}
\newcommand{\ee}{\end{equation}}  \newcommand{\eit}
{\end{itemize}}
\newcommand{\ba}{\begin{eqnarray}}\newcommand{\ea}{\end
{eqnarray}}

\begin{document}

\bc
\textbf{THE CONNECTION BETWEEN THE QUANTUM FREQUENCY 
OF RADIATION AND FREQUENCIES \\ 
OF CIRCLING OF THE ELECTRON IN ATOM OF HYDROGEN}
\ec
\medskip

\centerline{Stanislav Tereshchenko}
\centerline{Astrophysical Dept.}
\centerline{National Institut of Scientific and 
Technical Information}
\centerline{VINITI, Russian Academy of Sciences, Moscow}
\medskip

\begin{abstract}
The connection between the quantum frequency of 
radiation by the transition of the electron 
from orbit $n$ to orbit $k$ and frequencies of 
circling of electron in these orbits 
for the atom of hydrogen is determined.
\end{abstract}

In the atom of hydrogen by the transition of the 
electron from the orbit $n$ to 
the orbit $k$ ($n>k$) the quantum of energy is radiated
$$ h\nu = E_{n}-E_{k}\ .
$$
At present the dependence of quantum frequency of 
radiation $\nu $ from frequency 
of circling of electron $\nu _{n}$ and $\nu _{k}$ in 
orbit $n$ and $k$ is not determined. 
It is only known, that if the quantity of $n$ is big, 
by the transition $n\to{n-1}$ 
the quantum frequency $\nu $ is nearly equal to the 
frequency of circling of electron $\nu _{n}$ 
(principle of accordance).
However, one can show, that there is a simple 
connection between the quantum frequency 
of radiation $\nu $ and the frequencies of circling of 
electron $\nu _{n}$ and $\nu _{k}$. 
Indeed, the frequency of the circling of electron is 
equal to
$$
\nu _{n}= \frac{4\pi ^2me^4}{h^3n^3}\ .
$$
The quantum frequency of radiation is determined with 
the formula
\begin{equation} 
 \label{eq-1}
\nu =\frac{2\pi ^2me^4}{h^3}\left(\frac1{k^2}-\frac1
{n^2}\right)\ .
\end{equation}
The expression (\ref{eq-1}) we will written following 
way
$$
\nu = \frac{k}{2}\frac{4\pi ^2me^4}{h^3k^3}-\frac{n}{2}
\frac{4\pi ^2me^4}{h^3n^3}\ .
$$
Then
\begin{equation}
 \label{eq-2}
\nu =\frac12(k\nu _{k}-n\nu _{n})\ .
\end {equation}
Thus, we obtained the formula, which connects the 
quantum frequency of radiation $\nu$ 
with the frequencies circling of electron in atom 
hydrogen.
The product $n\nu _{n}$ makes the following sense: 
there are $n$ of  de Broglie's wave in the orbit $n$. 
Then $n\nu _{n}=\nu '_{n}$ is the frequency of de 
Broglie's wave of electron, 
moving in orbit $n$. So formula (\ref{eq-2}) can be 
written in following way
\begin{equation}
 \label{eq-3}
 \nu =\frac12(\nu'_{k}-\nu'_{n}) \ .
\end{equation}
It is possible, that formula (\ref{eq-3})  is also 
true for atoms with many electrons.

\end{document}